\documentclass[11pt]{article}
\usepackage{moriond,epsfig}

\def\be{\begin{equation}}
\def\ee{\end{equation}}
\def\bea{\begin{eqnarray}}
\def\eea{\end{eqnarray}}
\def\beqa{\begin{eqnarray}}
\def\eeqa{\end{eqnarray}}

\def\D0bar{\overline D{}^0}
\def\DDbar{D{}^0-\overline D{}^0}

\def\beq{\begin{equation}}
\def\eeq{\end{equation}}
\def\bea{\begin{eqnarray}}
\def\eea{\end{eqnarray}}

\newcommand{\DzDzb}{D^0-\overline{D^0}}

\begin{document}
\vspace*{4cm}
\title{CONSTRAINING NEW PHYSICS FROM $D^0-\overline{D}^0$ MIXING}

\author{ ALEXEY A. PETROV}

\address{Department of Physics and Astronomy, Wayne State University\\
Detroit, MI 48201, USA\\
Michigan Center for Theoretical Physics, University of Michigan\\
Ann Arbor, MI 48109, USA}

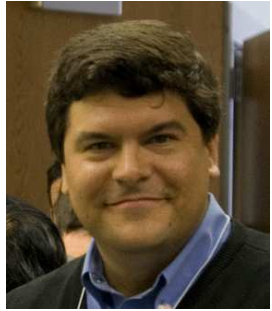
\maketitle\abstracts{
I review constraints on possible New Physics interactions from $D^0-\overline{D}^0$ mixing measurements. I consider 
the most general low energy effective Hamiltonian and include leading order QCD running of effective operators. 
I discuss constraints from an extensive list of popular New Physics models, each of which could be discovered at
the LHC,  that can generate these operators.  In most of the scenarios,  strong constraints that surpass those from 
other search techniques could be placed on the allowed parameter space using the existent evidence for observation 
of $D$ meson mixing. 
}

\section{Introduction}

Meson-antimeson mixing has traditionally been of importance 
because it is sensitive to heavy degrees of freedom that propagate in
the underlying mixing amplitudes. Estimates of the charm quark 
and top quark mass scales were inferred from the observation of mixing in 
the $K^0$ and $B_d$ systems, respectively, before these particles were 
discovered directly.

This success has motivated attempts to indirectly detect New Physics (NP) 
signals by comparing the observed meson mixing with predictions of 
the Standard Model (SM).  $K^0$-${\overline K}^0$ mixing has historically
placed stringent constraints on the parameter space of theories
beyond the SM and provides an essential hurdle that must be passed in the
construction of models with NP.  The large mixing signal in the $B_d$ and 
$B_s$ systems, observed at the B-factories and the Tevatron collider, can 
be precisely described in terms of the SM alone, which makes the parameter 
spaces of various NP models increasingly constrained. These facts
influenced theoretical and experimental studies of $D^0$ flavor 
oscillations, where the SM mixing rate is sufficiently small that the NP 
component might be able to compete. There has been a flurry of recent experimental 
activity regarding the detection of $D^0$-${\bar D}^0$ 
mixing, which marks the first time Flavor Changing Neutral Currents (FCNC)
have been observed in the charged $+2/3$ quark sector.  With the potential
window to discern large NP effects in the charm sector
and the anticipated improved accuracy for future mixing measurements,
the motivation for a comprehensive up-to-date theoretical analysis 
of New Physics contributions to $D$ meson mixing is compelling.

The phenomenon of meson-anti-meson mixing occurs in the presence of operators that
change quark flavor by two units~\cite{Artuso:2008vf}. Those operators can be generated 
in both the Standard Model and many possible extensions of it. They produce off-diagonal 
terms in the meson-anti-meson mass matrix, so that the basis of flavor eigenstates no longer 
coincide with the basis of mass eigenstates. Those two bases, however, are related by a linear 
transformation,
\beq
|D_{1\atop 2} \rangle = p | D^0 \rangle \pm q | \overline{D}^0 \rangle,
\eeq
where the complex parameters $p$ and $q$ are obtained from diagonalizing the 
$\DDbar$ mass matrix. Neglecting CP-violation leads to $p=q=1/\sqrt{2}$.
The mass and width splittings between those mass eigenstates are given by 
\beq\label{XandY}
x_D= \frac{m_1-m_2}{\Gamma_D}, \qquad y_D=\frac{\Gamma_1-\Gamma_2}{2 \Gamma_D}.
\eeq
It is expected that $x_D$ and $y_D$ should be rather small in the Standard Model, 
which is usually attributed to the absence of superheavy quarks destroying 
Glashow-Iliopoulos-Maiani (GIM) cancellation. In Eq.~(\ref{XandY}), $\Gamma_{\rm D}$ is 
the average width of the two neutral $D$ meson mass eigenstates.  The quantities which are 
actually measured in most experimental determinations of the mass and width differences, 
$y_{\rm D}^{\rm (CP)}$, $x_{\rm D}'$, and $y_{\rm D}'$, are defined as
\bea
y_{\rm D}^{\rm (CP)} &=&  y_{\rm D} \cos\phi - x_{\rm D}\sin\phi
\left(\frac{A_m}{2}-A_{prod}\right) \ \ , \nonumber \\
x_D' &=& x_D\cos\delta_{K\pi} + y_D\sin\delta_{K\pi} \ \ ,
\\
y_D' &=& y_{\rm D} \cos \delta_{K\pi} - x_{\rm D}
\sin\delta_{K\pi} \ \ ,
\nonumber
\label{y-defs}
\eea
where
$A_{prod} = \left(N_{D^0} - N_{{\overline D}^0}\right)/
\left(N_{D^0} + N_{{\overline D}^0}\right)$ is the so-called
production asymmetry of $D^0$ and $\overline{D}^0$ (giving
the relative weight of $D^0$ and ${\overline D}^0$ in the
sample) and $\delta_{K\pi}$ is the strong phase difference between
the Cabibbo favored and double Cabibbo suppressed
amplitudes~\cite{Bergmann:2000id}, which is usually measured in 
$D\to K\pi$ transitions. In what follows we shall neglect 
CP-violating parameters $\phi$ and $A_m$. In this limit 
$y_{\rm D}^{\rm (CP)}=y_D$. Please see recent 
reviews~\cite{Artuso:2008vf,Golowich:2008hu,Bianco:2003vb} for more complete analysis.

\section{Experimental Constraints on Charm Mixing}

The recent interest in $D^0$-${\bar D}^0$ mixing started with 
the almost simultaneous observations by the BaBar~\cite{Aubert:2007wf}
and Belle~\cite{Staric:2007dt} collaborations of nonzero mixing signals at about the per cent 
level,
\beqa
& & y_{\rm D}' = (0.97 \pm 0.44 \pm 0.31) \cdot 10^{-2} \qquad 
{\rm (BaBar)} \ , \\
& & y_{\rm D}^{\rm (CP)} = (1.31 \pm 0.32 \pm 0.25)\cdot 10^{-2} 
\qquad  {\rm (Belle)}\ \ .
\eeqa
This was soon followed by the announcement by the Belle collaboration 
of mixing measurements from the Dalitz plot analyses
of $D^0 \to K_S \pi^+ \pi^-$~\cite{Abe:2007rd}, 
\beq
x_{\rm D} = (0.80 \pm 0.29 \pm 0.17)\cdot 10^{-2}~, \ \ \qquad 
y_{\rm D} = (0.33 \pm 0.24 \pm 0.15)\cdot 10^{-2} \ \ .
\eeq
A fit to the current database by the Heavy Flavor Averaging Group (HFAG) 
gives~\cite{Schwartz:2008wa}
\beqa\label{hfag}
& & x_{\rm D} = 9.8^{+2.6}_{-2.7} \cdot 10^{-3}~, \quad 
y_{\rm D} = (7.5\pm 1.8) \cdot 10^{-3} \ \ ,
\eeqa
which is obtained assuming no CP-violation affecting mixing.
It is important to note that the combined analysis of $x_{\rm D}$ and $y_{\rm D}$
excludes the "no-mixing" point $x_{\rm D}=y_{\rm D}=0$ by $6.7\sigma$~\cite{Schwartz:2008wa}.
This fact adds confidence that charm mixing has indeed been observed. Then, a 
correct interpretation of the results is important. In addition, as with any rare low-energy
transition, the question arises on how to use it to probe for physics beyond the Standard Model.

\section{Standard Model "background" in $\DzDzb$ mixing}

Theoretical predictions for $x_D$ and $y_D$ obtained in the framework of the Standard Model 
historically span several orders of magnitude. I will not discuss predictions of the SM for 
the charm mixing rates here, instead referring the interested reader to recent 
reviews~\cite{Artuso:2008vf,Golowich:2008hu,Bianco:2003vb}. It might be advantageous to 
note that there are two approaches to describe $\DDbar$ mixing, neither of which give very 
reliable results because $m_c$ is in some sense intermediate between heavy and light. 

The ÒinclusiveÓ approach~\cite{Petrov:1997ch,Inclusive} is based on the operator product 
expansion (OPE). In the formal limit $m_c \gg \Lambda$ limit, where $\Lambda$ is a scale 
characteristic of the strong interactions, $x_D$ and $y_D$ can be expanded in terms of matrix 
elements of local operators. The use of the OPE relies on local quark- hadron duality, and on 
$\Lambda/m_c$ being small enough to allow a truncation of the series after the first few terms. 
This, however, is not realized in charm mixing, as the leading term in $1/m_c$ is suppressed by four and 
six powers of the strange quark mass for $x_D$ and $y_D$ respectively. The parametrically-suppressed 
higher order terms in $1/m_c$ can have less powers of $m_s$, thus being more important 
numerically~\cite{Inclusive}. This results in reshuffling of the OPE series, making it a triple expansion in
$1/m_c$, $m_s$, and $\alpha_s$. The (numerically) leading term contains over twenty matrix 
elements of dimension-12, eight-quark operators, which are difficult to compute reliably. A naive power 
counting then yields $x_D, y_D < 10^{-3}$. The ÒexclusiveÓ approach~\cite{Exclusive} sums over 
intermediate hadronic states. Since there are cancellations between states within a given $SU(3)$ 
multiplet, one needs to know the contribution of each state with high precision. However, the $D$ is not 
light enough that its decays are dominated by a few final states. In the absence of sufficiently precise data, 
one is forced to use some assumptions. Large effects in $y_D$ appear 
for decays close to $D$ threshold, where an analytic expansion in $SU(3)_F$ violation is no longer 
possible. Thus, even though theoretical calculations of $x_D$ and $y_D$ are quite uncertain, the values
$x_D \sim y_D \sim 1\%$ are quite natural in the Standard Model~\cite{Falk:2001hx}. 

It then appears that experimental results of Eq.~(\ref{hfag}) are consistent with the SM predictions. Yet, 
those predictions are quite uncertain to be subtracted from the experimental data to precisely
constrain possible NP contributions. In this situation the following approach can be taken. One can neglect the 
SM contribution altogether and assume that NP saturates the result reported by experimental collaborations. 
This way, however, only an upper bound on the NP parameters can be placed. A subtlety of this method of
constraining the NP component of the mixing amplitude is related to 
the fact that the SM and NP contributions can have either the same or opposite signs. While the sign of the SM 
contribution cannot be calculated reliably due to hadronic uncertainties, $x_D$ computed entirely within a given 
NP model can be determined rather precisely. This  stems from the fact that NP contributions are generated by 
heavy degrees of freedom making short-distance OPE reliable. This means that only the part of parameter 
space of NP models that generate $x_D$ of the same sign as observed experimentally can be reliably and 
unambiguously constrained.

\section{New Physics contributions to $\DzDzb$ mixing}

Any NP degree of freedom will generally be associated with a generic heavy mass scale $M$, 
at which the NP interaction will be most naturally described.  At the scale $m_c$ of the charm mass, 
this description will have been modified by the effects of QCD, which should be taken into account.  
In order to see how NP might affect the mixing amplitude, it is instructive to consider off-diagonal 
terms in the neutral D mass matrix,
\begin{eqnarray}\label{M12}
 \left (M - \frac{i}{2}\, \Gamma\right)_{12} =
 \frac{1}{2 M_{\rm D}} \langle \D0bar | 
{\cal H}_w^{\Delta C=-2} | D^0 \rangle 
+  \frac{1}{2 M_{\rm D}} \sum_n {\langle \D0bar | {\cal H}_w^{\Delta
  C=-1} | n \rangle\, \langle n | {\cal H}_w^{\Delta C=-1} 
| D^0 \rangle \over M_{\rm D}-E_n+i\epsilon}\,
\end{eqnarray}
where the first term contains ${\cal H}_w^{\Delta C=-2}$, which is an 
effective $|\Delta C| = 2$ hamiltonian, represented by a set of operators that are local at 
the $\mu \simeq m_D$ scale. Note that a $b$-quark also gives a (negligible) contribution
to this term. This term only affects $x_D$, but not $y_D$. 

The second term in Eq.~(\ref{M12}) is given by a double insertion of the effective $|\Delta C| = 1$ 
Hamiltonian ${\cal H}_w^{\Delta C=-1}$. This term is believed to give dominant contribution to 
$\DDbar$ mixing in the Standard Model, affecting both $x$ and $y$. It is generally believed that
NP cannot give any sizable contribution to this term, since ${\cal H}_w^{\Delta C=-1}$ Hamiltonian 
also mediates non-leptonic $D$-decays, which should then also be affected by this NP contribution.
I will show that there is a well-defined theoretical limit where NP contribution {\it dominates} 
lifetime difference $y_D$ and consider implications of this limit in "real world".

\subsection{New Phyiscs in $|\Delta C|=1$ interactions.}  

Consider a non-leptonic $D^0$ decay amplitude, $A[D^0 \to n]$, which includes a small NP contribution, 
$A[D^0 \to n]=A_n^{\rm (SM)} + A_n^{\rm (NP)}$. Here, $A_n^{\rm (NP)}$ is assumed to be smaller than
the current experimental uncertainties on those decay rates. This ensures that NP effects cannot be seen
in the current experimental analyses of non-leptonic D-decays. One can then write $y_{\rm D}$ as
\begin{eqnarray}\label{schematic}
y_{\rm D} &\simeq& \sum_n \frac{\rho_n}{\Gamma_{\rm D}} 
A_n^{\rm (SM)} A_n^{\rm (SM)}
+ 2\sum_n \frac{\rho_n}{\Gamma_{\rm D}}
A_n^{\rm (NP)} A_n^{\rm (SM)} \ \ . 
\end{eqnarray}
The first term of Eq.~({schematic}) represents the SM contribution to $y_{\rm D}$. The SM contribution to $y_{\rm D}$ is 
known to vanish in the limit of exact flavor $SU(3)$. Moreover, the first order correction is also absent, 
so the SM contribution arises only as a {\it second} order effect~\cite{Falk:2001hx}. This means that
in the flavor $SU(3)$ limit the lifetime difference $y_{\rm D}$ is dominated by the second term in 
Eq.~(\ref{schematic}), i.e. New Physics contributions, even if their contibutions are tiny in the individual decay 
amplitudes~\cite{Golowich:2006gq}!  A calculation reveals that NP contribution to $y_{\rm D}$ can be as large 
as several percent in R-parity-violating SUSY models~\cite{Petrov:2007gp} or as small as $\sim 10^{-10}$ 
in the models with interactions mediated by charged Higgs particles~\cite{Golowich:2006gq}.

This wide range of theoretical predictions can be explained by two observations. First, many NP 
affecting $|\Delta C| = 1$ transitions also affect  $|\Delta B| = 1$ or $|\Delta S| = 1$ decays or
kaon and B-meson mixings, which are tightly constrained. Second, a detailed look at a given NP model 
that can potentially affect $y_{\rm D}$ reveals that the NP contribution itself can vanish in the flavor $SU(3)$ limit.
For instance, the structure of the NP interaction might simply mimic the one of the SM. Effects like that 
can occur in some models with extra space dimensions. Also, the chiral structure of a low-energy effective 
lagrangian in a particular NP model could be such that the leading, mass-independent contribution vanishes 
exactly, as in a left-right model (LRM). Finally, the NP coupling might explicitly depend on the quark mass, 
as in a model with multiple Higgs doublets. However, most of these models feature second order 
$SU(3)$-breaking already at leading order in the $1/m_c$ expansion. This should be contrasted with the SM, 
where the leading order is suppressed by six powers of $m_s$ and term of order $m_s^2$ only appear as a 
$1/m_c^6$-order correction.

\subsection{New Phyiscs in $|\Delta C|=2$ interactions.}

Though the particles present in models with New Physics may not be produced in charm quark decays, 
their effects can nonetheless be seen in the form of effective operators generated by the exchanges
of these new particles. Even without specifying the form of these new interactions, we know that 
their effect is to introduce several $|\Delta C|=2$ effective operators built out of the SM degrees of freedom.

By integrating out new degrees of freedom associated with new interactions at a scale $M$, we are
left with an effective hamiltonian written in the form of a series of operators of increasing dimension. 
Operator power counting then tells us the most important contributions are given by the operators of the 
lowest possible dimension, $d=6$ in this case. This means that they must contain only
quark degrees of freedom and no derivatives. Realizing this, we can write the complete basis of
these effective operators, which can be done most conveniently in terms of chiral quark fields,
\beq\label{SeriesOfOperators}
\langle f | {\cal H}_{NP} | i \rangle =
G \sum_{i=1} {\rm C}_i (\mu) ~
\langle f | Q_i  | i \rangle (\mu) \ \ ,
\eeq
where the prefactor $G$ has the dimension of inverse-squared mass, the 
${\rm C}_i$ are dimensionless Wilson coefficients, and the $Q_i$ are the effective operators:
\beqa
\begin{array}{l}
Q_1 = (\overline{u}_L \gamma_\mu c_L) \ (\overline{u}_L \gamma^\mu
c_L)\ , \\
Q_2 = (\overline{u}_L \gamma_\mu c_L) \ (\overline{u}_R \gamma^\mu
c_R)\ , \\
Q_3 = (\overline{u}_L c_R) \ (\overline{u}_R c_L) \ , \\
Q_4 = (\overline{u}_R c_L) \ (\overline{u}_R c_L) \ ,
\end{array}
\qquad
\begin{array}{l}
Q_5 = (\overline{u}_R \sigma_{\mu\nu} c_L) \ ( \overline{u}_R
\sigma^{\mu\nu} c_L)\ , \\
Q_6 = (\overline{u}_R \gamma_\mu c_R) \ (\overline{u}_R \gamma^\mu
c_R)\ , \\
Q_7 = (\overline{u}_L c_R) \ (\overline{u}_L c_R) \ , \\
Q_8 = (\overline{u}_L \sigma_{\mu\nu} c_R) \ (\overline{u}_L
\sigma^{\mu\nu} c_R)\ \ .
\end{array}
\label{SetOfOperators}
\eeqa
In total, there are eight possible operator structures that exhaust the
list of possible independent contributions to $|\Delta C|=2$ transitions.
Since these operators are generated at the scale $M$ where the New Physics is 
integrated out, a non-trivial operator mixing can occur when one takes into account
renormalization group running of these operators between the scales $M$
and $\mu$, with $\mu$ being the scale where the hadronic matrix elements are computed.
We shall work at the renormalization scale $\mu= m_c \simeq 1.3$~GeV.
This evolution is determined by solving the RG equations obeyed by
the Wilson coefficients,
\beq\label{AnomEq}
\frac{d}{d \log \mu} \vec C (\mu) = \hat \gamma^T \vec C (\mu)\ \ ,
\eeq
where $\hat \gamma$ represents the matrix of anomalous dimensions of
the operators in Eq.~(\ref{SetOfOperators})~\cite{Golowich:2007ka}. 
Due to the relatively simple structure of $\hat\gamma$,
one can easily write the evolution of each Wilson coefficient in
Eq.~(\ref{SeriesOfOperators}) from the New Physics scale
${\rm M}$ down to the hadronic scale $\mu$, taking into
account quark thresholds.  Corresponding to each of the eight
operators $\{ Q_i \}$ ($i=1,\dots,8$) is an RG factor
$r_i (\mu , M)$.  The first of these, $r_1(\mu,M)$, is given
explicitly by
\beqa
r_1(\mu,M)&=& \left(\frac{\alpha_s(M)}{\alpha_s(m_t)}\right)^{2/7}
\left(\frac{\alpha_s(m_t)}{\alpha_s(m_b)}\right)^{6/23}
\left(\frac{\alpha_s(m_b)}{\alpha_s(\mu)}\right)^{6/25} \ \ .
\label{wilson}
\eeqa
and the rest can be expressed in terms of $r_1(\mu,M)$ as
\beqa
\begin{array}{l}
r_2(\mu,M) = [r_1(\mu,M)]^{1/2} \ , \\
r_3(\mu,M) = [r_1(\mu,M)]^{-4} \ , \\
r_4(\mu,M) = [r_1(\mu,M)]^{(1 + \sqrt{241})/6} \ ,\\
r_5(\mu,M) = [r_1(\mu,M)]^{(1 - \sqrt{241})/6} \ ,
\end{array}
\qquad
\begin{array}{l}
\phantom{xxxx} \\
r_6(\mu,M) = r_1(\mu,M) \ , \\
r_7(\mu,M) = r_4(\mu,M) \ , \\
r_8(\mu,M) = r_5(\mu,M) \ \ . 
\end{array}
\eeqa
The RG factors are generally only weakly dependent on the NP scale
$M$ since it is taken to be larger than the top quark mass, $m_t$,
and the evolution of $\alpha_s$ is slow at these high mass scales.
In Table~\ref{tab:LO}, we display numerical values for the
$r_i (\mu , M)$ with $M = 1,2$~TeV and $\mu = m_c \simeq 1.3$~GeV.
Here, we compute $\alpha_s$ using the one-loop evolution and matching 
expressions for perturbative consistency with the RG evolution of the
effective hamiltonian.
\begin{table}[t]
\begin{center}
\begin{tabular}{|c||c|c|c|c|c|}
\hline
$M{\rm (TeV)}$ & $r_1(m_c, M)$ & $r_2(m_c, M)$ & $r_3(m_c, M)$ 
& $r_4(m_c, M)$ & $r_5(m_c, M)$ \\
\hline \hline
$1$ & $0.72$ & $0.85$ & $3.7$ & $0.41$ & $2.2$ \\
\hline
$2$ & $0.71$ & $0.84$ & $4.0$ & $0.39$ & $2.3$ \\
\hline
\end{tabular}
\end{center}
\vskip .05in\noindent
\caption{Dependence of the RG factors on the heavy mass scale $M$.}
\label{tab:LO}
\end{table}
A contribution to $\DDbar$ mixing from a 
particular NP model can be obtained by calculating matching conditions for the 
Wilson coefficients $C_i$ at the scale $M$, running their values down to $\mu$ and
computing the relevant matrix elements of four-quark operators. A generic model of 
New Physics would then give the following contribution $x_{\rm D}$, 
\begin{eqnarray}\label{GenericNP}
x_{\rm D}^{NP} = G ~{f_D^2B_Dm_D\over
\Gamma_D}  \left[ {2\over 3}[C_1(m_c)+C_6(m_c)] - {5\over 12}
[C_4(m_c)+C_7(m_c])+ {7\over 12}C_3(m_c) \right. \nonumber\\
\left. -{5C_2(m_c)\over
6}+[C_5(m_c)+C_8(m_c)] \right]\ \ .
\end{eqnarray}
Here we simplified the result by assuming that all non-perturbative ('bag') parameters 
are equal to $B_D\simeq 0.82$. The Wilson coefficients at the scale $\mu$
are related to the Wilson coefficients at the scale $M$ by renormalization group
evolution,
\begin{eqnarray}
C_1(m_c) & = & r_1(m_c, M)C_1(M)\ ,\nonumber\\
C_2(m_c) & = & r_2(m_c, M)C_2(M)\ ,\nonumber\\
C_3(m_c) & = & {2\over 3}\left[ r_2(m_c, M)-
r_3(m_c, M)\right] C_2(M) +
r_3(m_c, M)C_3(M) \ ,\nonumber\\
C_4(m_c) & = & {8\over\sqrt{241}}\left[ r_5(m_c, M)- 
r_4(m_c, M)\right]\left[C_4(M) +{15\over 4}
C_5(M)\right]\nonumber\\
& & \hspace{0.5cm} + {1\over 2}\left[ r_4(m_c, M)+
r_5(m_c, M)\right]C_4(M) \ ,\nonumber\\
C_5(m_c) & = & {1\over 8\sqrt{241}}\left[ r_4(m_c, M)- 
r_5(m_c, M)\right]\left[C_4(M)+ 64C_5(M)
\right]\nonumber\\
& & \hspace{0.5cm} + {1\over 2}\left[ r_4(m_c, M)+
r_5(m_c, M)\right]C_5(M)\ ,\nonumber\\
C_6(m_c) & = & r_6(m_c, M)C_6(M)\ ,\\ 
C_7(m_c) & = &  {8\over\sqrt{241}}\left[ r_8(m_c, M)- 
r_7(m_c, M)\right]\left[C_7(M) +{15\over 4}
C_8(M)\right]\nonumber\\
& & \hspace{0.5cm} + {1\over 2}\left[ r_7(m_c, M)+
r_8(m_c, M)\right]C_7(M) \ ,\nonumber\\
C_8(m_c) & = & {1\over 8\sqrt{241}}\left[ r_7(m_c, M)- 
r_8(m_c, M)\right]\left[C_7(M)+ 64C_8(M)
\right]\nonumber\\
& &  \hspace{0.5cm}  + {1\over 2}\left[ r_7(m_c, M)+
r_8(m_c, M)\right]C_8(M)\ ,\nonumber 
\end{eqnarray}
A contribution of each particular NP model can then be studied using Eq.~(\ref{GenericNP}).
Even before performing such an analysis, one can get some idea what energy 
scales can be probed by $\DDbar$ mixing. Setting $G=1/M^2$ and $C_i(M)=1$, we
obtain $M \sim 10^3$~TeV. More realistic models can be probed in the region of several TeV, 
which is very relevant for LHC phenomenology applications. 

A program described above has been recently executed~\cite{Golowich:2007ka} for 21 
well-motivated NP models, which will be actively studied at LHC. The results are presented in 
Table~\ref{tab:bigtableofresults}.
As can be seen, out of 21 models considered, only four received no useful constraints
from $\DDbar$ mixing. More informative exclusion plots can be found in that 
paper~\cite{Golowich:2007ka} as well. It is interesting to note that some models 
{\it require} large signals in the charm system if mixing and FCNCs in the strange 
and beauty systems are to be small (as in, for example, the SUSY alignment 
model~\cite{Nir:1993mx,Ciuchini:2007cw}). 
\begin{table}[t]
\begin{center}
\begin{tabular}{|c||c|}
\hline 
Model & Approximate Constraint 
\\ \hline\hline
Fourth Generation  &\ \ $|V_{ub'} V_{cb'}|\cdot m_{b'}  
<   0.5 $~(GeV)  
\ \ \\
$Q=-1/3$ Singlet Quark  
&  $s_2\cdot m_S  < 0.27$~(GeV) 
\\
$Q=+2/3$ Singlet Quark 
&  $|\lambda_{uc}| < 2.4 \cdot 10^{-4}$ \\
Little Higgs  &  Tree: See entry for $Q=-1/3$ Singlet Quark \\
& Box: Parameter space can reach 
observed $x_{\rm D}$
\\
Generic $Z'$
&  $M_{Z'}/C > 2.2\cdot 10^3$~TeV  \\
Family Symmetries  & $m_1/f>1.2\cdot 10^{3}$~TeV 
 (with $m_1/ m_2 = 0.5$)   \\
Left-Right Symmetric   & No constraint   \\
Alternate Left-Right Symmetric  &
$M_R>1.2$~TeV ($m_{D_1}=0.5$~TeV)   \\
 & ($\Delta m/m_{D_1})/M_R>0.4$~TeV$^{-1}$\\
Vector Leptoquark Bosons 
& $M_{VLQ} > 55 (\lambda_{PP}/0.1) $~TeV  \\
Flavor Conserving Two-Higgs-Doublet 
&   No constraint \\
Flavor Changing Neutral Higgs 
&  $m_H/C>2.4\cdot 10^3$~TeV \\
FC Neutral Higgs (Cheng-Sher)  & 
$m_H/|\Delta_{uc}|>600$~GeV\\
Scalar Leptoquark Bosons  & See entry for RPV SUSY \\
Higgsless   & $M > 100$~TeV  \\
Universal Extra Dimensions & No constraint \\
Split Fermion  
& $M / |\Delta y| > (6 \cdot 10^2~{\rm GeV})$
 \\
Warped Geometries &  $M_1 > 3.5$~TeV \\
MSSM  & 
$|(\delta^u_{12})_{\rm LR,RL}| 
< 3.5 \cdot 10^{-2}$ for ${\tilde m}\sim 1$~TeV  \\
  & 
$|(\delta^u_{12})_{\rm LL,RR}| < .25 $ for ${\tilde m}\sim 1$~TeV  
\\
SUSY Alignment & ${\tilde m} > 2$~TeV  \\
Supersymmetry with RPV & 
$\lambda'_{12k} \lambda'_{11k}/m_{\tilde d_{R,k}} < 
1.8 \cdot 10^{-3}/100$~GeV
\\
Split Supersymmetry & No constraint  \\
\hline\hline
\end{tabular}
\end{center}
\vskip .05in\noindent
\caption{Approximate constraints on NP models from $D^0$ mixing.}
\label{tab:bigtableofresults}
\end{table}
%

\section{Conclusions}

I reviewed implications of recent measurement of $\DDbar$ mixing rates for
constraining models of New Physics. A majority of considered models received 
competitive constraints from $\DDbar$ mixing measurements despite hadronic 
uncertainties that plague SM contributions. It should be noted that vast majority of
predictions of NP models do not suffer from this uncertainty, and can be
computed reliably, if lattice QCD community provides calculations of
matrix elements of four-fermion operators Eq.~(\ref{SetOfOperators}).

Another possible manifestation of new physics interactions in the charm
system is associated with the observation of (large) 
CP-violation~\cite{Artuso:2008vf,Bianco:2003vb,Petrov:2007ms}. 
This is due to the fact that all quarks that build up the hadronic states in weak 
decays of charm mesons belong to the first two generations. Since $2\times2$ 
Cabbibo quark mixing matrix is real, no CP-violation is possible in the
dominant tree-level diagrams which describe the decay amplitudes. 
CP-violating amplitudes can be introduced in the Standard Model by including 
penguin or box operators induced by virtual $b$-quarks. However, their 
contributions are strongly suppressed by the small combination of 
CKM matrix elements $V_{cb}V^*_{ub}$. It is thus widely believed that the 
observation of (large) CP violation in charm decays or mixing would be an 
unambiguous sign for New Physics.

\section*{Acknowledgments}

This work was supported in part by the U.S.\ National Science Foundation
CAREER Award PHY--0547794, and by the U.S.\ Department of Energy under Contract
DE-FG02-96ER41005.


\end{document}